\documentclass[12pt]{iopart}

\usepackage{graphicx}
\usepackage{subfigure}
\usepackage{color}

\begin{document}

\title[Mechanical losses of silicon flexures at low temperatures]{Investigation of mechanical losses of thin silicon flexures at low temperatures}

\author{R. Nawrodt$^{1,2}$, C. Schwarz$^2$, S. Kroker$^{2,3}$, I. W. Martin$^1$, F. Br\"uckner$^3$, L. Cunningham$^1$, V. Gro{\ss}e$^2$, A. Grib$^4$, D. Heinert$^2$, J. Hough$^1$, T. K\"asebier$^3$, E. B. Kley$^3$, R. Neubert$^2$, S. Reid$^1$, S. Rowan$^1$, P. Seidel$^2$, M. Th\"urk$^2$, A. T\"unnermann$^{3,5}$}

\address{$^1$Institute for Gravitational Research, University of Glasgow, G12 8QQ, United Kingdom}
\address{$^2$Institut f\"ur Festk\"orperphysik, Friedrich-Schiller-Universit\"at Jena, Helmholtzweg 5, D-07743 Jena, Germany}
\address{$^3$Institut f\"ur Angewandte Physik, Friedrich-Schiller-Universit\"at Jena, Albert-Einstein-Stra{\ss}e 15, D-07745 
Jena, Germany}
\address{$^4$Kharkiv V. N. Karazin National University, Physics Department, 61077 Kharkiv, Ukraine}
\address{$^5$Fraunhofer-Institut f\"ur Angewandte Optik und Feinmechanik IOF, Albert-Einstein-Stra{\ss}e 7, D-07745 Jena, Germany}

\ead{r.nawrodt@physics.gla.ac.uk}

\begin{abstract}

The investigation of the mechanical loss of different silicon flexures in a temperature region from 5 to 300\,K is presented. The flexures have been prepared by different fabrication techniques. A lowest mechanical loss of $3\times10^{-8}$ was observed for a 130\,$\mu$m thick flexure at around 10\,K. While the mechanical loss follows the thermoelastic predictions down to 50\,K a difference can be observed at lower temperatures for different surface treatments. This surface loss will be limiting for all applications using silicon based oscillators at low temperatures. The extraction of a surface loss parameter using different results from our measurements and other references is presented. We focused on structures that are relevant for gravitational wave detectors. The surface loss parameter $\alpha_s$ = 0.5\,pm was obtained. This reveals that the surface loss of silicon is significantly lower than the surface loss of fused silica.

\end{abstract}

%\submitto{\CQG}
%\maketitle

\section{Introduction}

Gravitational wave (GW) detectors currently under operation are amongst the most sensitive instruments ever built sensitive for displacements. They are measuring tiny fluctuations of the space-time based on an interferometric principle. The optical components of a Michelson-like interferometer are suspended as pendula in order to reduce seismic coupling. Current detectors like LIGO \cite{Abramovici1992}, Virgo \cite{Bernardini1999} or TAMA \cite{Takamori2002} use metal wire to suspend the test masses. GEO600 \cite{Plissi1998} pioneered the use of a monolithic suspension system based on fused silica fibres \cite{Braginsky1994,Rowan1997} and hydroxide catalysis bonding \cite{Gwo2001,Rowan1998}. Second generation detectors like Advanced LIGO \cite{AdvLIGO} or Advanced Virgo \cite{Flaminio2005} will adopt this technique among others and aim for an increase of sensitivity of about an order of magnitude compared to initial detectors. Currently, a design study for a 3$^{rd}$ generation detector is underway in Europe \cite{ETProposal}. In order to further increase the sensitivity beyond the 2$^{nd}$ generation detectors a dramatic change in size, topology and materials for the detector are needed (see e.g. \cite{Hild2010}). Silicon is a promising candidate material for such a GW observatory due to its excellent thermal and mechanical properties \cite{Winkler1991, Rowan2005}. Silicon would allow the fabrication of large optical substrates as well as suspension elements in a quasi-monolithic suspension \cite{Alshourbagy2006,Veggel2009}. Furthermore, first attempts have been made to demonstrate a full reflective coating based on monolithic waveguides \cite{Brueckner2008}. 

The mechanical loss of a material is a critical parameter to estimate the thermal noise performance of an optical component or a suspension (see e.g. \cite{Nawrodt2009et}). However, crystalline materials have in general a larger coefficient of thermal expansion compared to amorphous materials like fused silica. This results in a large thermoelastic noise at room temperature \cite{Braginsky1999}. The use of crystalline materials like silicon makes it necessary to operate the detector at low temperatures to reduce the contribution of thermoelastic noise. 

It has been shown in the past that the mechanical loss of small structures can be deconvoluted into a bulk material intrinsic loss and a surface loss \cite{Gretarsson1999,Yasumura2000,Yang2002}. This surface loss can be a significant and limiting source of mechanical loss -- especially in small scale structures like suspension elements or the micro-structured surfaces of optical elements like monolithic waveguides. Our investigation focuses on structures which sizes that are relevant for the construction of suspensions and optical micro and nano structures. Furthermore, we exclusively focused on surface preparation techniques that are applicable for a long time operation in vacuum.

%We present measurements of the mechanical loss of silicon flexures with geometries that are suitable for a suspension element of a 3$^{rd}$ generation GW detector. The experimental investigation covers a frequency band from 1 to 86\,kHz and a temperature region from 5 to 300\,K. By means of extracting the mechanical loss values at low temperatures it is possible to extract the intrinsic and surface losses. A well accepted theory of surface loss in fused silica \cite{Gretarsson1999} is used to explain the behavior of silicon at low temperatures. The characteristic surface loss parameter is extracted and compared to fused silica.

\section{Sample preparation and experimental technique}

The mechanical loss of silicon flexures was measured using a ring-down technique described elsewhere in detail \cite{Reid2006,Martin2008}. The flexures were fabricated from high purity Si(100) wafers with their long axis parallel to the Si(110) direction. All flexures have a thicker block (525\,$\mu$m) on one side and a thin flexure (75-130\,$\mu$m) with lengths between 35 and 50\,mm. The thin flexures were obtained by two techniques: wet chemical etching and dry chemical etching. When the thin flexure is oscillating the thicker block at one end will only contain a reduced amount of vibrational energy allowing the flexures to be clamped at this side without affecting the mechanical loss significantly. Table\,{\ref{tab:samples}} summarizes the parameters of the samples used in this paper.

\begin{table}
\caption{\label{tab:samples}Summary of the parameters of the Si flexures.}
\begin{indented}
\item[]\begin{tabular}{@{}lllll}
\br
Properties & Sample 1 & Sample 2 & Sample 3 & Sample 4\\
\mr
length (mm) & 35\,$\pm$\,0.2 & 50\,$\pm$\,0.2 & 50\,$\pm$\,0.2 & 50\,$\pm$\,0.2\\
width (mm) & 4\,$\pm$\,0.2 & 8\,$\pm$\,0.3 & 8\,$\pm$\,0.3 & 8\,$\pm$\,0.3\\
thickness ($\mu$m) &  &  & & \\
\hspace{2mm} flexure & 75\,$\pm$\,2 & 100\,$\pm$\,5 & 130\,$\pm$\,10 & 130\,$\pm$\,5\\
\hspace{2mm} block   & 525 & 525 & 525 & 525\\
etching & wet & dry & dry & dry\\ 
roughness &   &   &   &   \\
\hspace{2mm} front & 33\,nm & 6.8\,nm & 5.9\,nm & 6.2\,nm \\
\hspace{2mm} back  & 0.9\,nm& 1.9\,nm & 1.6\,nm & 337\,nm\\
\br
\end{tabular}
\end{indented}
\end{table}

\begin{figure}
%\begin{indented}
%\item[]
\begin{center}
\subfigure[]{\includegraphics[scale=0.33]{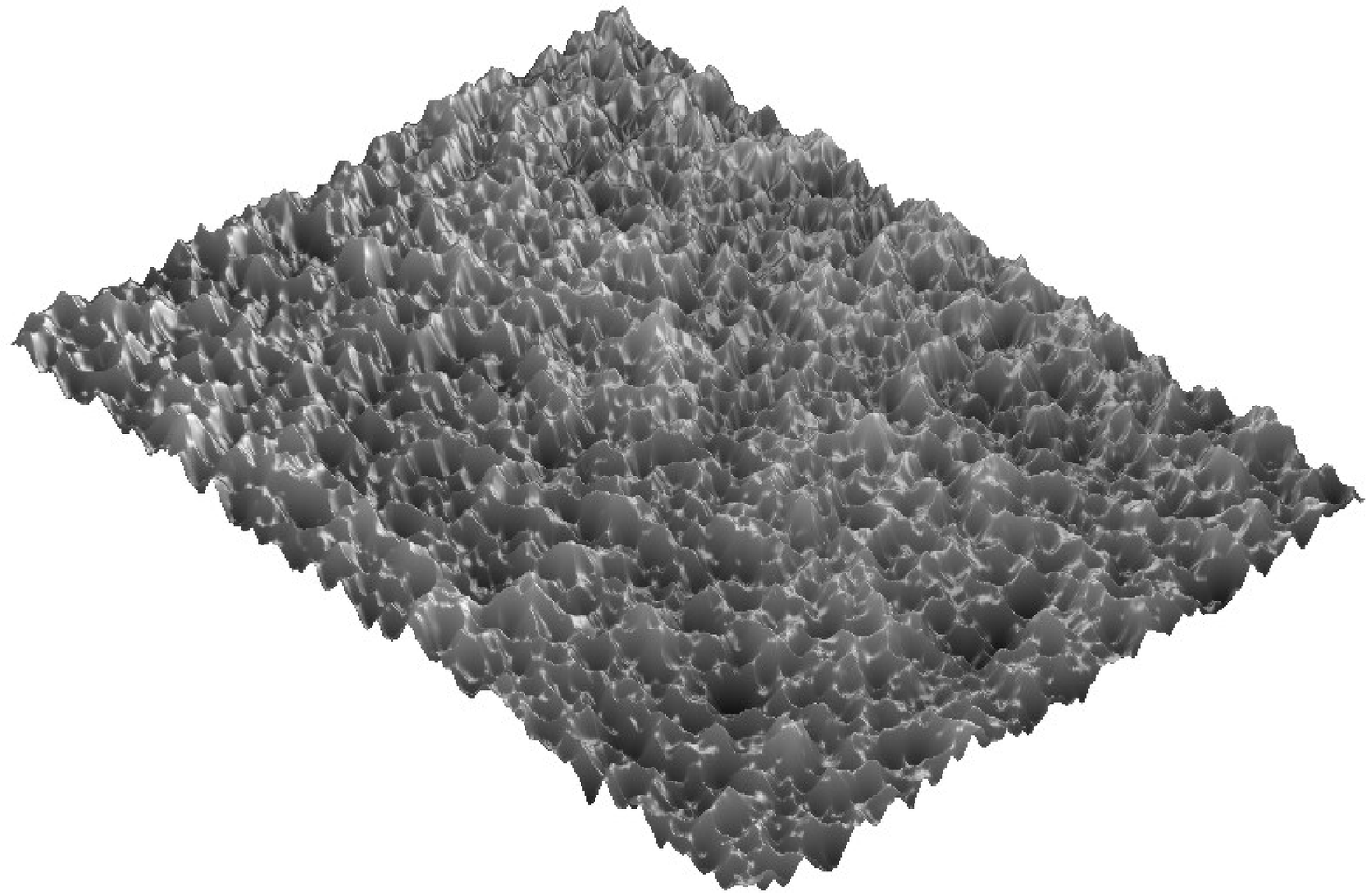}}
\subfigure[]{\includegraphics[scale=0.33]{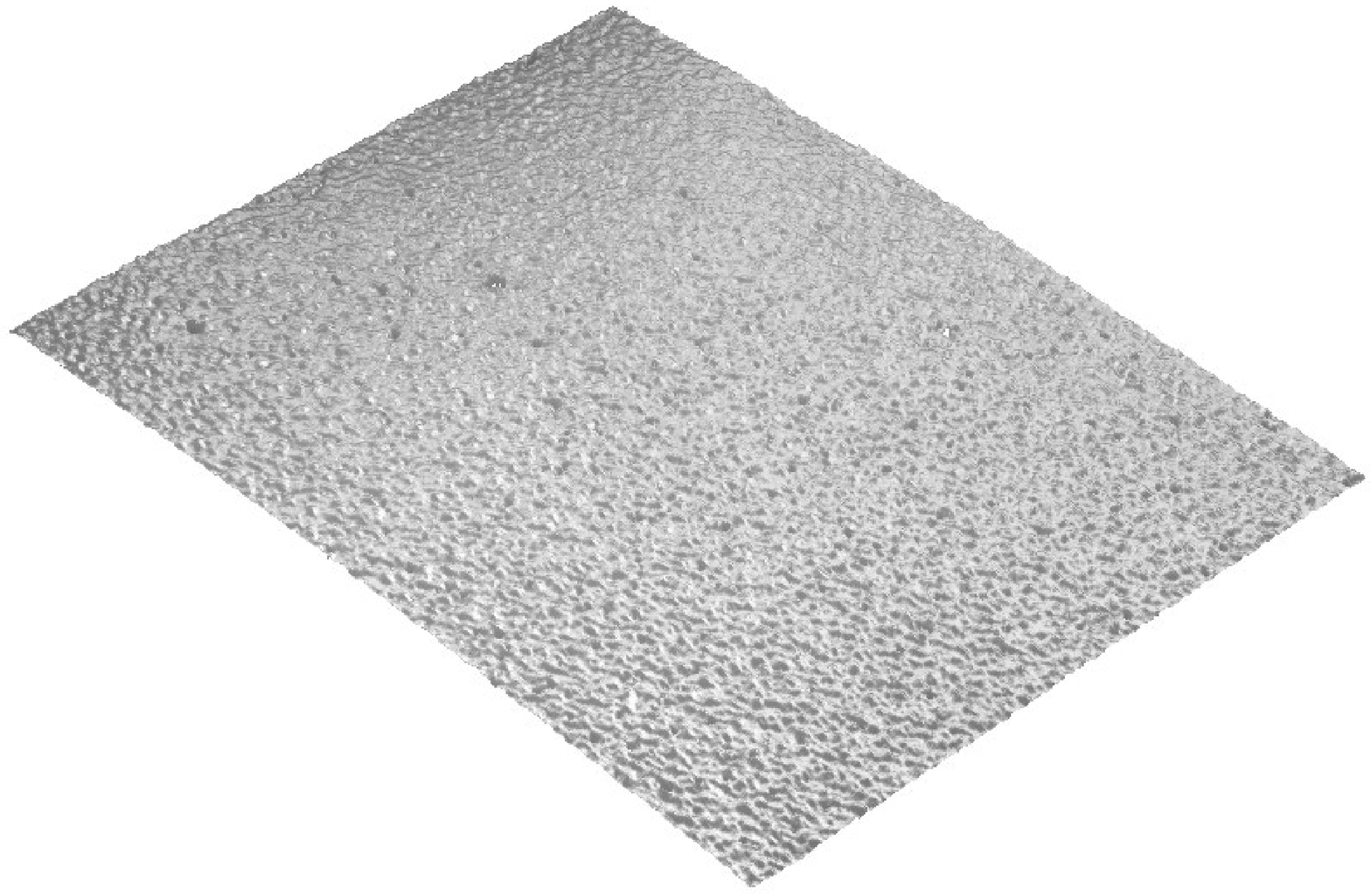}}
\end{center}
\caption{\label{fig:surface}Surface profile of a wet etched (a) and a dry etched (b) sample. Both images have been obtained with the same magnification and show an area of 600\,$\mu m$ $\times$ 450\,$\mu m$.}
%\end{indented}
\end{figure}

Due to the different fabrication techniques the surface topology of the samples is also different (see figure\,\ref{fig:surface}). The wet chemical etching shows etch pits. The dry etched samples have a smooth surface with small ($\mu$m scale) droplets. The roughness of the dry etched samples is about a factor of 5 smaller than for the wet chemical etched. Sample 4 was prepared from a single side polished wafer. The etching process was carried out from the polished side. The result was a silicon flexure with an identical geometry to sample 3 but one unpolished side. This sample was used to investigate the influence of the surface on the measured mechanical loss.

Internal resonant modes of the samples were excited by means of an electrostatic drive plate and an applied high voltage (up to 1600\,V). The excitation using an appropriate driving frequency allows the selection of different modes. The resonant frequencies $f_n$ of the flexures can be calculated in the case of bending modes by means of \cite{McLachlan1951}:

\begin{equation}
f_n = \frac{\alpha_n^2t}{2\pi L^2}\sqrt{\frac{Y}{12\rho}}
\label{eq:res_freq}
\end{equation}

\noindent $t$ is the thickness of the flexure, $L$ its length, $Y$ the Young's modulus and $\rho$ the density of the material. $\alpha_n$ is a numerical factor and can be obtained from:

\begin{equation}
\cos \alpha_n \times \cosh \alpha_n = -1
\end{equation}

\noindent The first four values are 1.875, 4.694, 7.855 and 10.996.

The vibration of the silicon flexure was observed by a simple readout which senses the motion of a laser beam reflected from the surface. When a sufficiently large amplitude was reached the excitation was switched off and the subsequent ring down was recorded. The characteristic ring down time $\tau$ (which corresponds to the amplitude decay to 1/e of the initial value) leads immediately to the mechanical loss $\phi$:

\begin{equation}
\phi = \frac{1}{\pi f_0 \tau}
\end{equation}

\noindent with the resonant frequency $f_0$ of the mode under investigation. The whole setup was placed into a specially built cryostat \cite{Nawrodt2006} providing the desired temperature and its stability over the measuring period. At low temperatures a typical temperature stability of better than 0.3\,K was achieved within the ring down period. Typical ring down times $\tau$ in the order of several 10 up to 1000\,s have been observed for frequencies between 1 and 86\,kHz. All modes have been measured several times during 3 temperature cycles from 5 to 300\,K. Each sample was re-clamped twice to reduce systematic errors arising from bad clamping.

\section{Mechanical loss in silicon flexures}

The measured loss of a silicon flexure will consist of different types of dissipation that causes energy loss. Being an integral measurement it is important to carefully study possible loss sources in order to be able to interpret the mechanical loss spectra throughout the full frequency and temperature region. The most likely types of mechanical dissipation occurring in the presented experiment are thermoelastic losses, phonon-phonon scattering losses, losses that are related to surface states and impurities or other losses that shall be summarized as excess losses. 

\subsection{Thermoelastic loss}

Zener \cite{Zener1938} was the first to investigate a loss mechanism due to dissipative heat flux in samples under bending oscillation. When a sample is bent parts are locally heated and cooled due to the inversion of the thermal expansion law. These local temperature differences cause heat fluxes that are accompanied by an increase of entropy -- or dissipation. The temperature distribution $T$ in an object can be obtained from the stress field $\sigma_{ij}$ by means of the generalized equation of heat conductivity \cite{Lifshitz2000,Norris2005}:

\begin{equation}
\rho C_p\dot{T} - \kappa_{ij} \frac{\partial T}{\partial x_i \partial x_j} = -\alpha_{ij}\dot{\sigma_{ij}} T_0
\label{eq:gen_heatcond}
\end{equation}

\noindent $\kappa_{ij}$ and $\alpha_{ij}$ are the tensors of thermal conductivity and thermal expansion. $T_0$ is the average temperature of the sample. $C_p$ is the heat capacity. The mechanical loss induced by thermoelastic damping is then given by:

\begin{equation}
\phi_{TE} = \frac{1}{2E} \int_V \alpha_{ij} \hat{\sigma_{ij}} Im\{\hat{T} \} dV
\label{eq:numerical_loss}
\end{equation}

\noindent $E$ is the total vibrational energy stored in the test body, $T = \hat{T} e^{i\omega t}$ and $\sigma = \hat{\sigma} e^{i\omega t}$ with $\omega = 2\pi f$ the angular frequency of the mode under consideration. Eq.\,(\ref{eq:gen_heatcond}) and eq.\,(\ref{eq:numerical_loss}) can be solved by means of a finite element analysis for arbitrarily shaped test bodies and arbitrary mode shape.

Eq.\,(\ref{eq:numerical_loss}) can be derived analytically in the special case of pure bending modes in isotropic materials. For these modes the restoring force for the oscillation is exclusively based on bending forces. The simplified equation is given by \cite{Zener1938}:

\begin{equation}
\phi_{TE} = \frac{\alpha^2YT}{\rho C_p}\frac{\omega\tau}{1+\omega^2\tau^2}
\label{eq:TE}
\end{equation}

\noindent with

\begin{equation}
\tau = \frac{\rho C_pt^2}{\pi \kappa}
\end{equation}

\noindent where $t$ is the thickness of the sample.

\subsection{Phonon-phonon loss}

Another possible loss source in silicon is the phonon-phonon loss. In thermal equilibrium the phonon distribution is defined by the temperature $T$ of the sample. Applying an external oscillation with a typical wavelength much longer than the phonon wavelength (like throughout this paper) results in a modulation of the lattice. This modulation shifts the phonon distribution. The process of reassembling all phonons to this new local equilibrium generates entropy and is thus a loss mechanism which is called Akhiezer loss. The mechanical loss associated with this mechanism is given by \cite{Boemmel1960}:

\begin{equation}
\phi_{ph-ph}=\frac{C_pT\gamma^2}{v^2}\frac{\omega\tau_{ph}}{1+\omega^2\tau_{ph}^2}
\end{equation}

\noindent with the parameter $\gamma=2.2$, the speed of sound $v$ and the mean phonon lifetime $\tau_{ph}$ which can be obtained from the thermal conductivity \cite{Boemmel1960}. This assumption is not correct at very low temperatures where the thermal conductivity is limited by the sample dimensions \cite{Casimir1938}. However, this treatment will allow a comparison of the different contributions.

\subsection{Surface loss}

Several authors have pointed out that a mechanical loss contribution of a micro cantilever is dominated by a thin surface layer (see e.g. \cite{Gretarsson1999,Yasumura2000,Yang2002,Wang2003,Liu2005}). This area is assumed to have different mechanical parameters compared to the bulk values. The origin of this change in parameters can be manifold, like local lattice distortions, adsorbed materials at the surface, open bonds, surface roughness, etc, and is not fully understood. The surface loss contribution can be written as (see e.g. \cite{Gretarsson1999}):

\begin{equation}
\label{eq:surf_loss}
\phi = \phi_{bulk} \left(1+\mu \frac{d_s}{V/S} \right)
\end{equation}

\noindent with the mechanical loss of the bulk material $\phi_{bulk}$ and the volume-to-surface ratio $V/S$. The dissipation depth $d_s$ can be written as:

\begin{equation}
\label{eq:dissdepth}
d_s = \frac{1}{\phi_{bulk}Y_{bulk}} \int_0^h \phi(n) Y(n) dn.
\end{equation}

\noindent Here, $Y_{bulk}$ is the bulk's Young's modulus, $\phi(n)$ and $Y(n)$ are the distributions of the mechanical loss and the Young's modulus within the thin surface layer. $h$ is the thickness of the surface layer. This description already assumes that the inhomogeneity of the surface layer is only dependent on the depth $n$ from the surface.

The numerical factor $\mu$ in eq.\,(\ref{eq:surf_loss}) takes the geometry of the sample and the modeshape of the resonance under investigation into account. It is given by:

\begin{equation}
\label{eq:mu}
\mu = \frac{V}{S}\frac{\int\int_S \epsilon^2(\vec{r}) d^2r}{\int\int\int_V \epsilon^2(\vec{r})d^3r}.
\end{equation}

\noindent $\epsilon$ is the strain amplitude and V and S are the volume and the surface area. For a flexure with rectangular cross-section oscillating in a transverse vibration this leads to \cite{Gretarsson2000}:

\begin{equation}
\label{eq:mu_simple}
\mu = \frac{3+t/w}{1+t/w}
\end{equation}

\noindent with the thickness $t$ and width $w$ of the flexure. Thin flexures lead to a constant value of $\mu=3$. In the general case eq.\,(\ref{eq:mu}) can be solved numerically using a FEA program to estimate the strain amplitudes $\epsilon$.

It is often convenient to write eq.\,(\ref{eq:surf_loss}) in a different form:

\begin{equation}
\label{eq:surf_loss_simple}
\phi = \phi_{bulk} + \alpha_s \mu \frac{S}{V}
\end{equation}

\noindent with the surface loss parameter $\alpha_s=\phi_{bulk} d_s$.

A similar approach for the surface loss was made by Yasumura et al. \cite{Yasumura2000}. Assuming a thin flexure they obtained for the mechanical Q-factor of a surface loss dominated sample:

\begin{equation}
\label{eq:yasumura}
Q_{surface} = \frac{t}{6\delta}\frac{Y_{bulk}}{Y_{s}} Q_{s}.
\end{equation}

\noindent $Q_{s}$ is the mechanical Q-factor of the surface layer (which is the reciprocal of the mechanical loss), $t$ the thickness of the flexure, $\delta$ the thickness of the lossy surface layer and $Y_{s}$ the Young's modulus of the surface layer. Assuming that the surface layer has a similar Young's modulus as the bulk material and rewriting eq.\,(\ref{eq:yasumura}) in terms of losses leads to:

\begin{equation}
\phi_{surface} = \delta \phi_s \frac{6}{t}.
\end{equation}

\noindent This coincides with the more general eq.\,(\ref{eq:surf_loss}) in the case of thin flexures ($\mu = 3, V/S \approx 2/t$) and the limit that the surface losses dominate the bulk losses. The surface loss parameter $\alpha_s$ can then be identified with the product $\phi_s \delta$.

\subsection{Excess loss}

All losses so far are directly linked to the sample being investigated. Additional loss mechanisms occur during the interaction of the sample with the measuring setup. These additional losses are called excess loss. This term is a summary of different types of loss like residual gas damping (see e.g. \cite{Braginsky1986,Blom1992}) or electrostatic losses from the driving plate \cite{Braginsky1986,Mitrofanov2000}. These losses can be reduced by carefully setting up the experiment and have been checked before the measurements \cite{Reid2006}. 

\subsection{Summary of possible internal loss mechanisms}

\begin{figure}[b!]
\begin{indented}
\item[]
%\begin{center}
\includegraphics[scale=0.7]{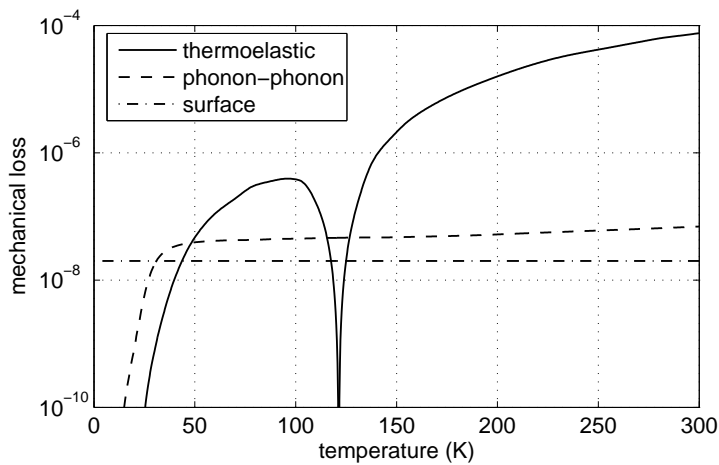}
%\end{center}
\caption{\label{fig:sum_loss}Summary of possible intrinsic loss sources of a silicon flexure at 10\,kHz. A surface loss of $2\times10^{-8}$ was assumed for the plot.}
\end{indented}
\end{figure}

Figure\,\ref{fig:sum_loss} summarizes the internal mechanical loss contributions for a typical flexure under investigation in this paper (sample 3 from table\,\ref{tab:samples}) for a frequency of 10\,kHz. The material properties have been obtained from standard databases \cite{Hull1999}.

Thermoelastic loss dominates at temperatures above approximately 50\,K. At around 125\,K the effect of thermoelastic loss is significantly reduced due to the vanishing coefficient of thermal expansion $\alpha$ of silicon at this temperature. Here, the phonon-phonon loss sets the achievable limit. 

Depending on the exact value of the surface loss it will dominate at low temperatures. In the intermediate temperature region between 25 and 50\,K the Akhiezer damping is the dominating loss source. Thus, in a setup with neglectable excess loss it should be possible to investigate the surface loss at temperatures below 20\,K (see section \ref{sec:surf_loss}).

\section{Experimental results}
\subsection{Mechanical loss at low temperatures}

\begin{figure}[h!]
%\begin{indented}
%\item[]
\begin{center}
\subfigure[]{\includegraphics[scale=0.52]{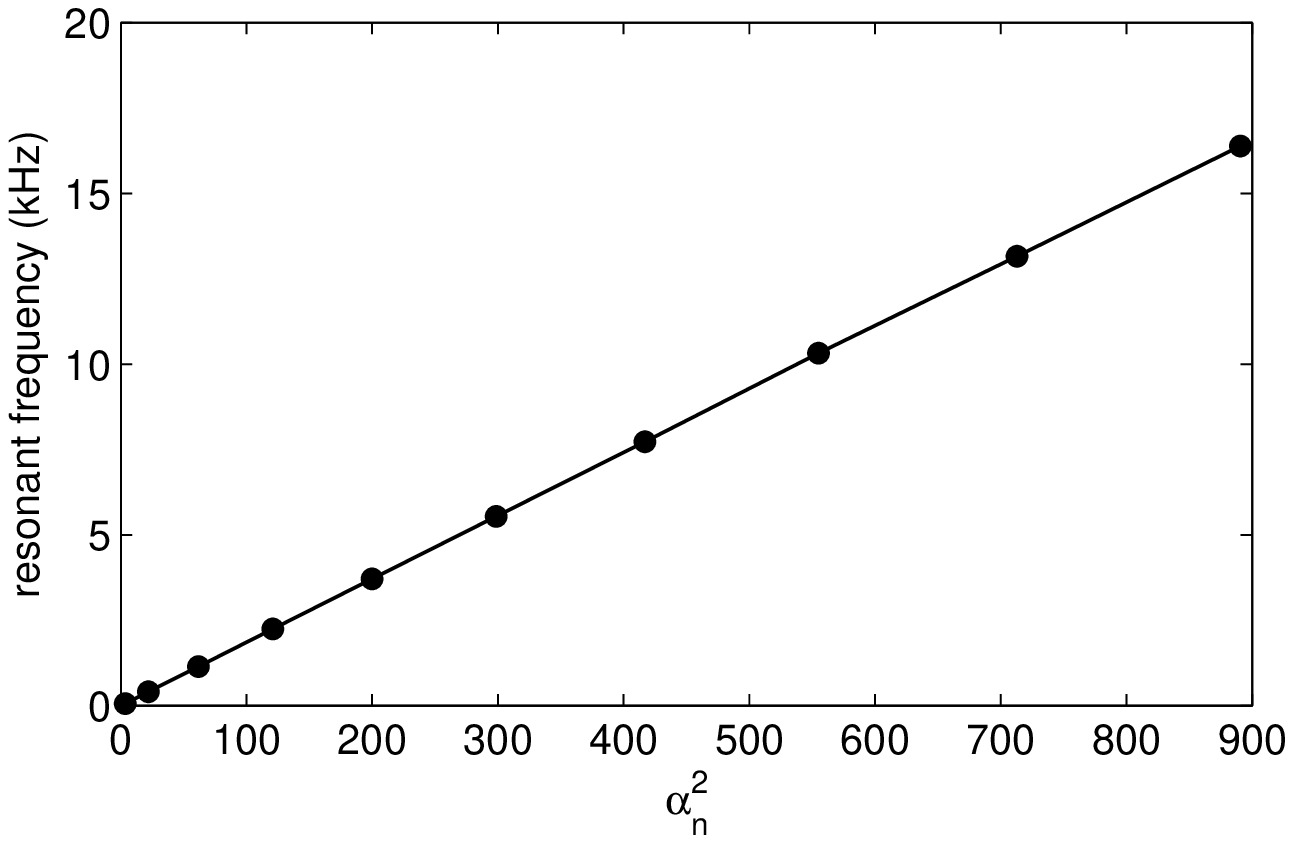}}
\subfigure[]{\includegraphics[scale=0.52]{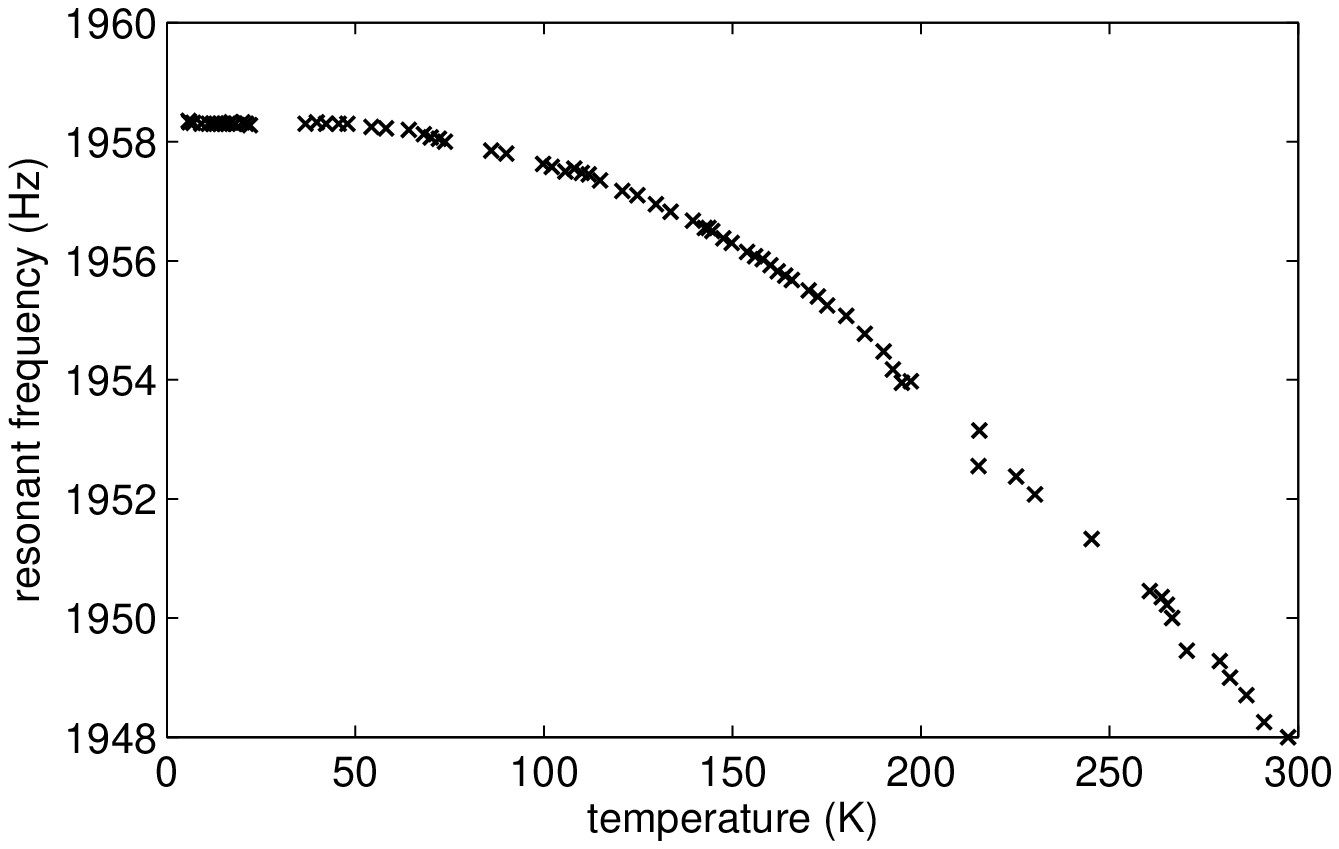}}
\end{center}
\caption{\label{fig:res_exp}(a) - Plot of the determined resonant frequency at 300\,K against $\alpha_n^2$ for the first 10 modes of sample 1. The linear fit represents the behavior predicted by eq.\,(\ref{eq:res_freq}) indicating that only bending modes have been selected. (b) - Temperature dependency of a typical resonant mode (3$^{rd}$ bending mode of sample 2). The temperature dependency is caused by the temperature dependent elastic constants. }
%\end{indented}
\end{figure}

For samples 1 to 3 several modes have been measured. Starting with the fundamental bending mode and using eq.\,(\ref{eq:res_freq}) it is possible to find the bending mode frequencies. It was possible to follow the sequence of bending modes up to about 70\,kHz. Additional modes resulted in a total number of resonant frequencies per run of more than 30. The higher frequency modes have only been measured at lower temperatures. Close to room temperature their mechanical losses are high and the corresponding ring down times too short to be measured with sufficient accuracy. Figure\,\ref{fig:res_exp} (a) shows the dependence of the measured resonant frequency on the parameter $\alpha_n$. Plotting $f_n$ against $\alpha_n^2$ should follow a straight line as predicted by eq.\,(\ref{eq:res_freq}). The fit can be used to determine the thickness of the samples. 

The resonant frequency is temperature dependent. This behavior is determined by the temperature dependence of the Young's modulus (more exactly: by the temperature dependence of the elastic constants). At temperatures below 50\,K the resonant frequency is approximately temperature independent. All frequencies throughout this paper are thus given at low temperatures unless otherwise noted.

\begin{figure}
%\begin{indented}
%\item[]
\begin{center}
\subfigure[3727\,Hz]{\includegraphics[scale=0.52]{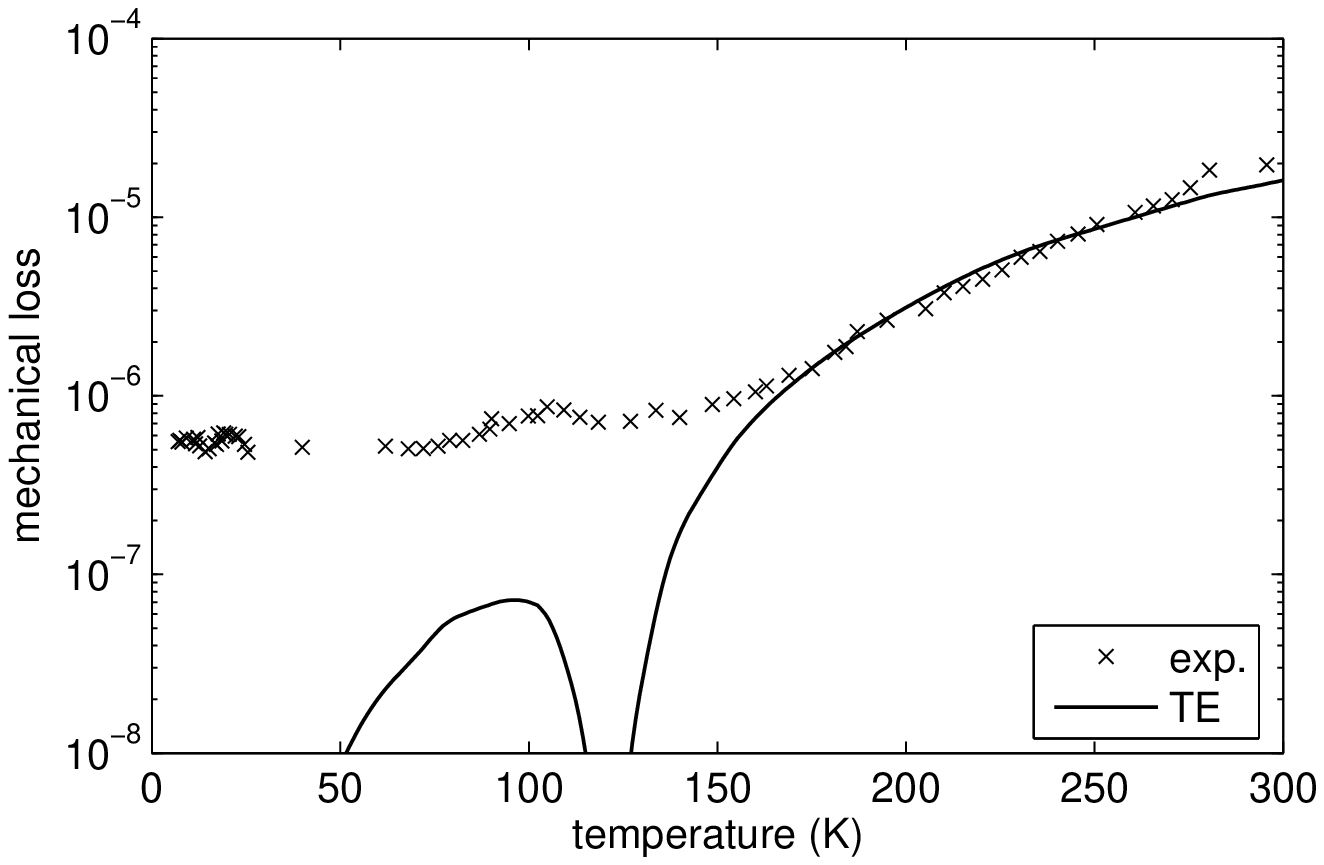}}
\subfigure[13225\,Hz]{\includegraphics[scale=0.52]{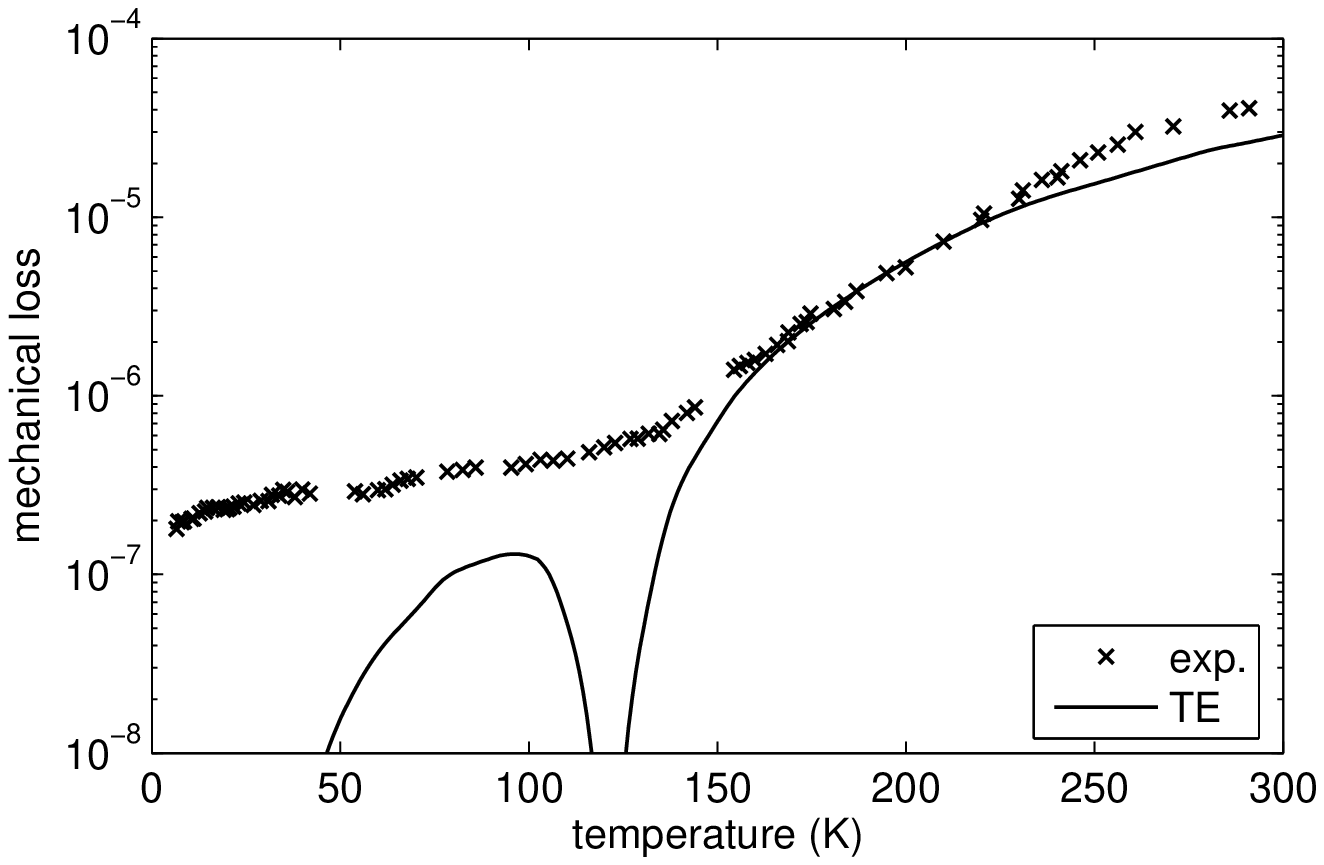}}
\subfigure[19980\,Hz]{\includegraphics[scale=0.52]{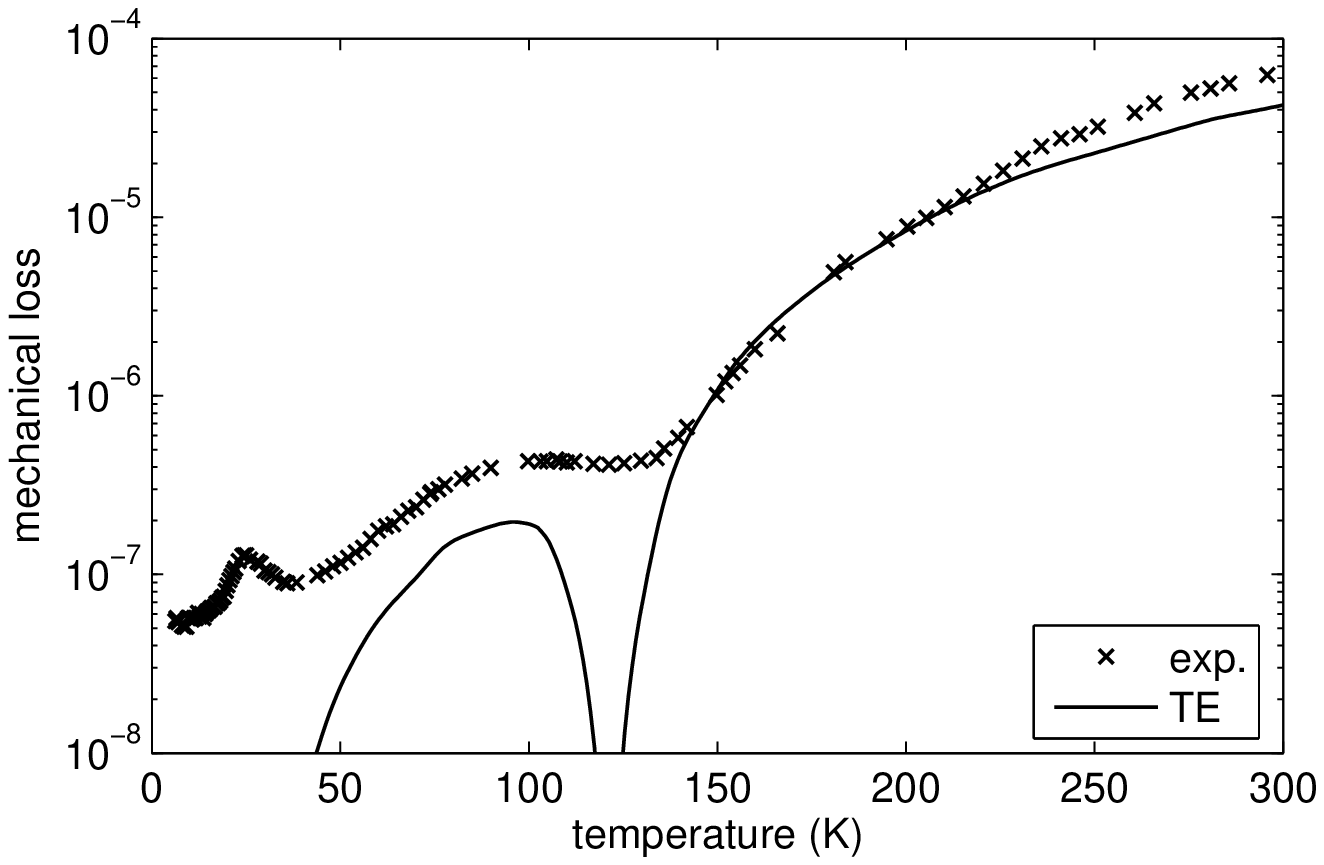}}
\subfigure[28236\,Hz]{\includegraphics[scale=0.52]{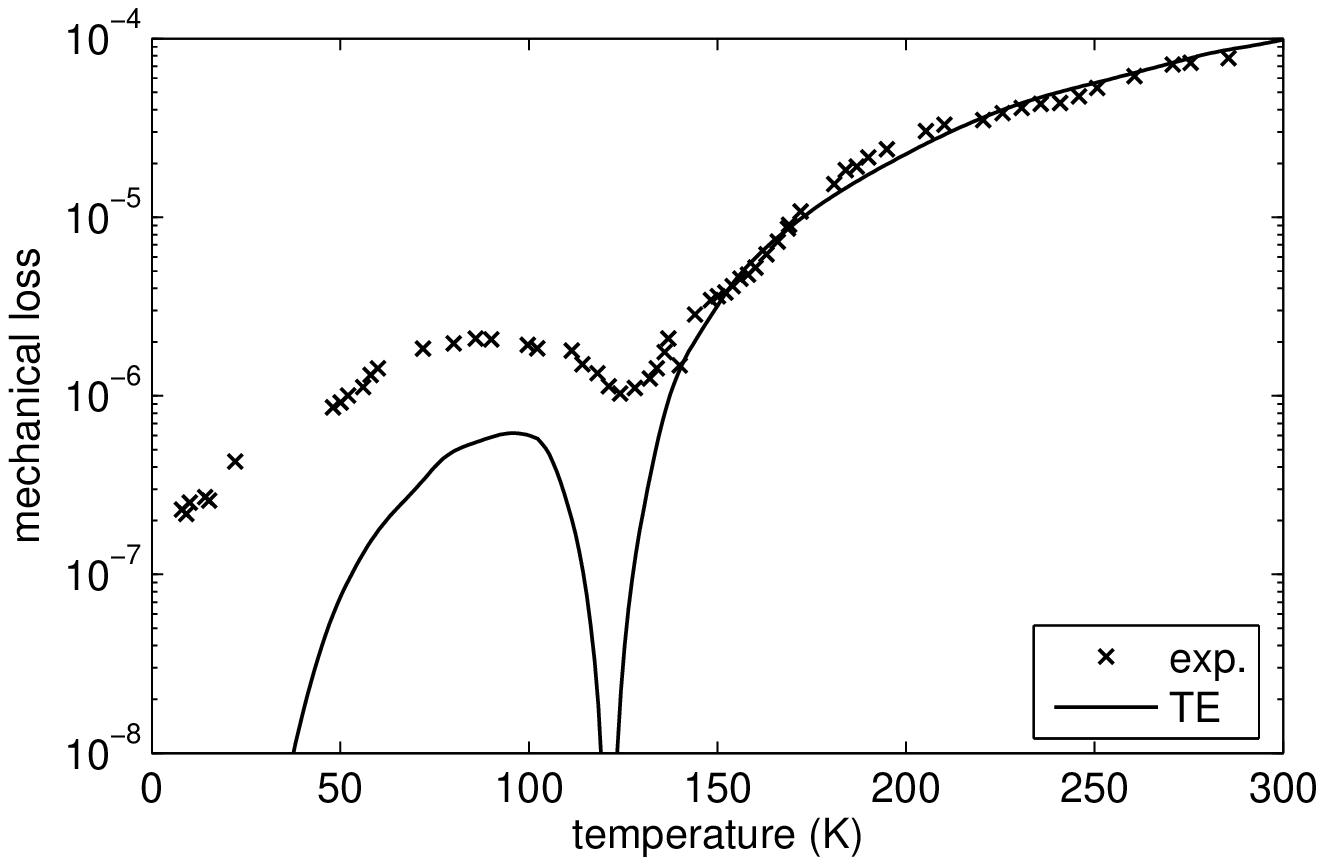}}
\end{center}
\caption{\label{fig:S1}Experimental results of the mechanical loss of sample 1 (wet etching, size).}
%\end{indented}
\end{figure}

\begin{figure}
%\begin{indented}
%\item[]
\begin{center}
\subfigure[1958\,Hz]{\includegraphics[scale=0.52]{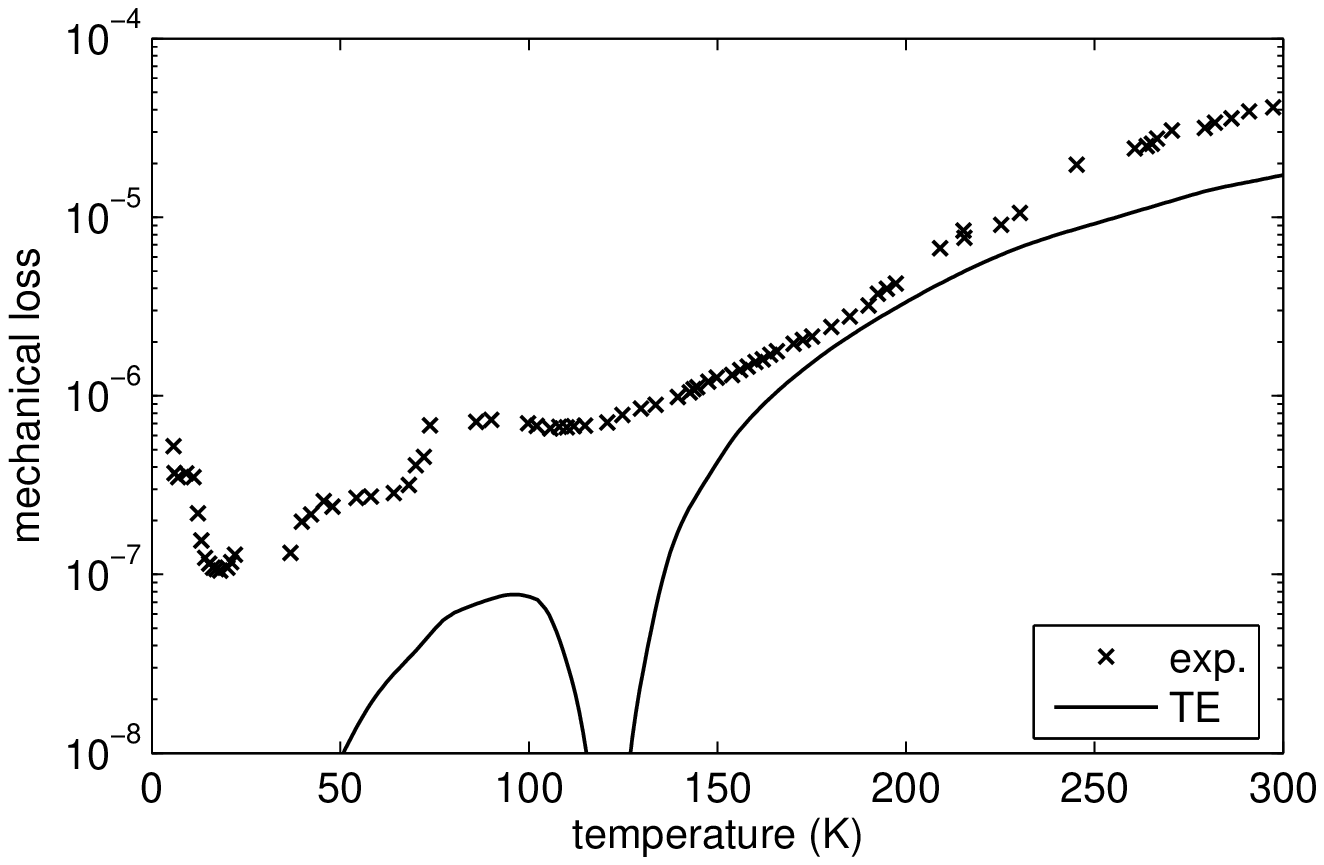}}
\subfigure[14592\,Hz]{\includegraphics[scale=0.52]{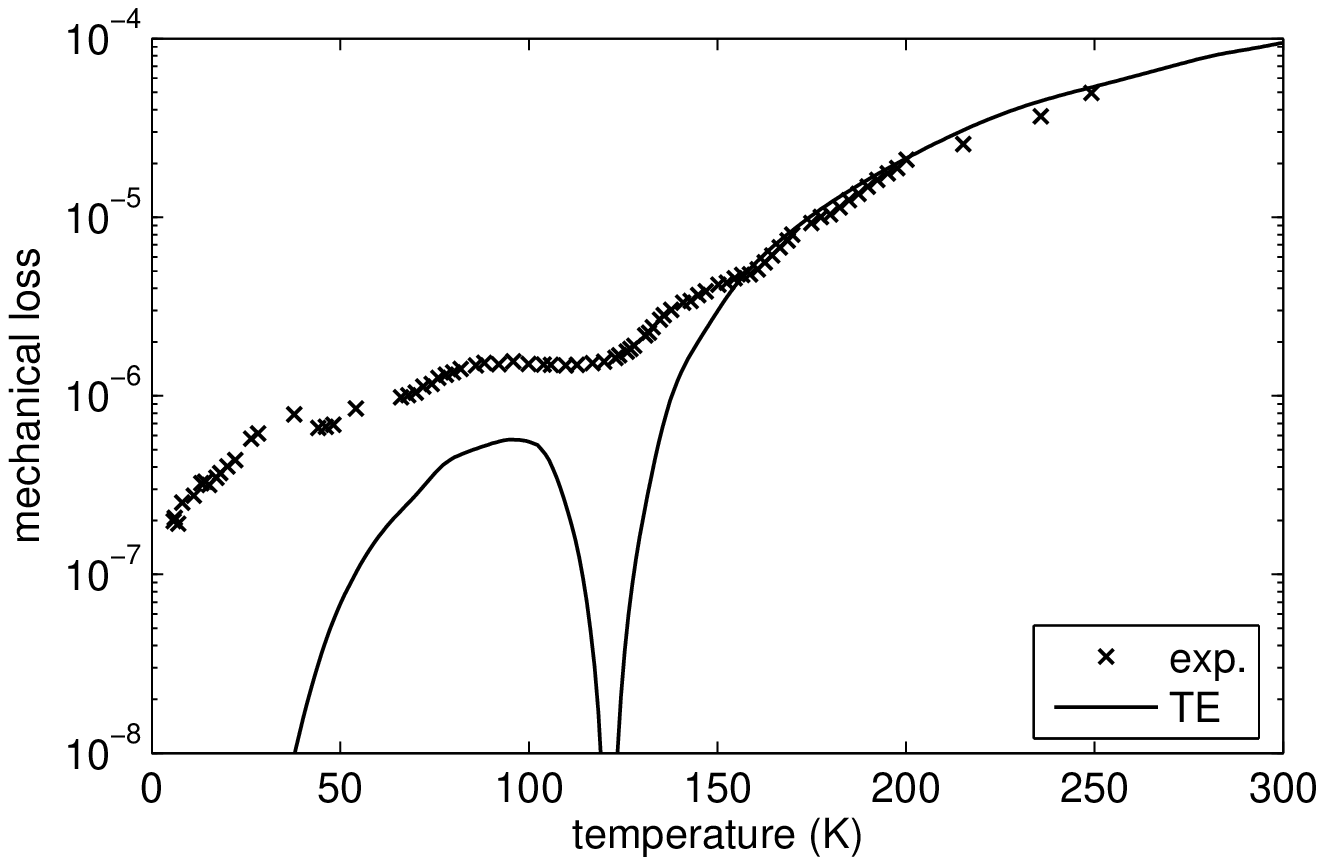}}
\subfigure[18922\,Hz]{\includegraphics[scale=0.52]{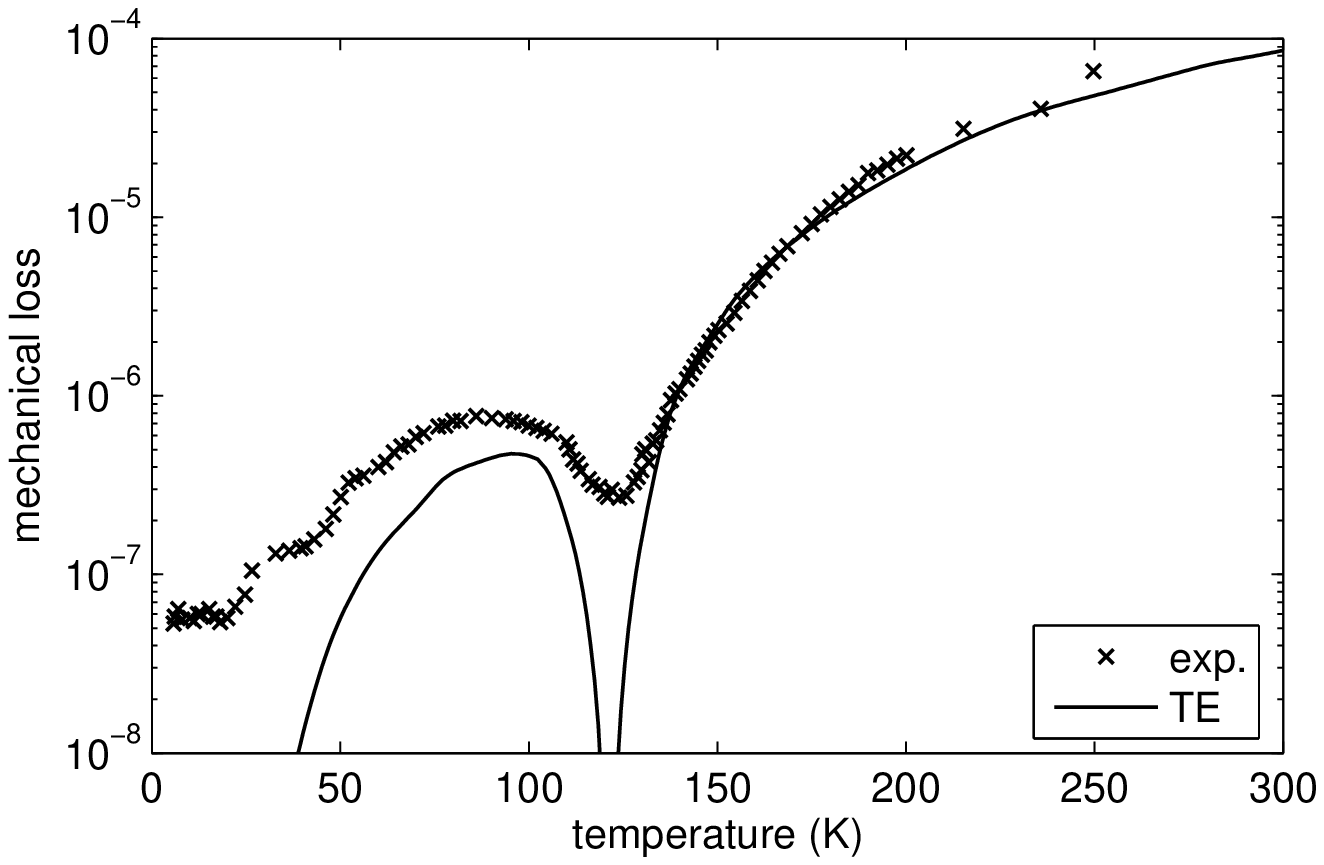}}
\subfigure[86587\,Hz]{\includegraphics[scale=0.52]{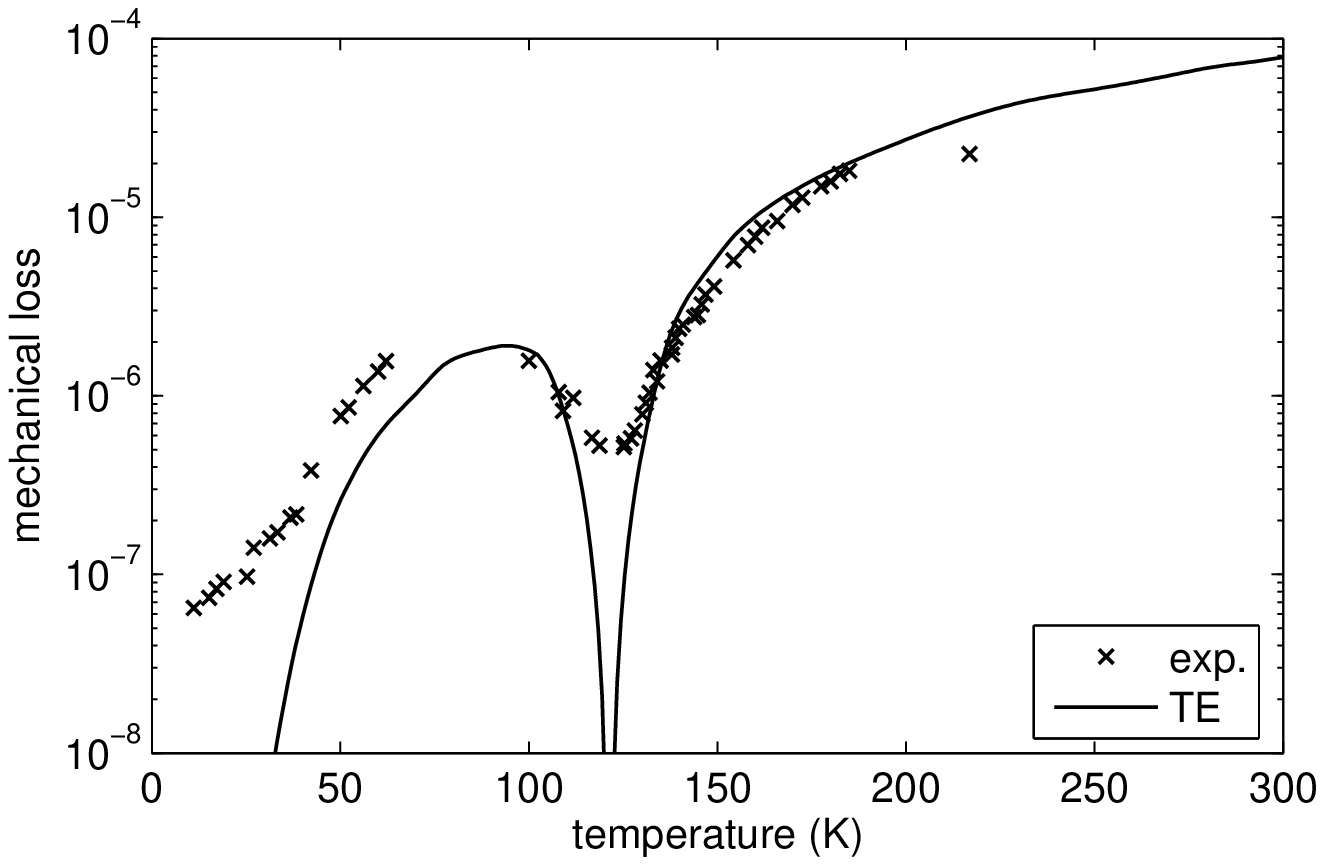}}
\end{center}
\caption{\label{fig:S2}Experimental results of the mechanical loss of sample 2 (dry etching, size).}
%\end{indented}
\end{figure}

\begin{figure}[b!]
%\begin{indented}
%\item[]
\begin{center}
\subfigure[1201\,Hz]{\includegraphics[scale=0.52]{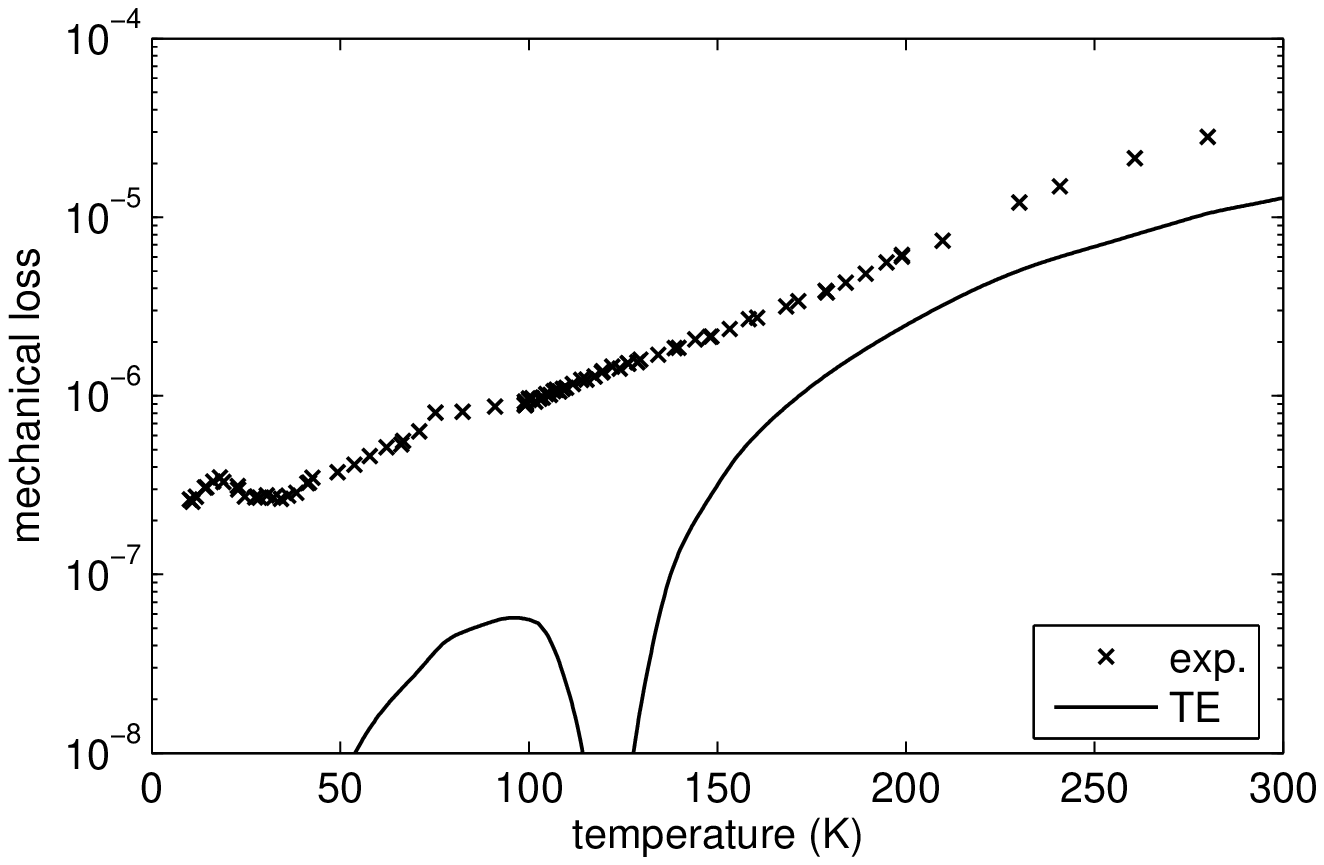}}
\subfigure[18370\,Hz]{\includegraphics[scale=0.52]{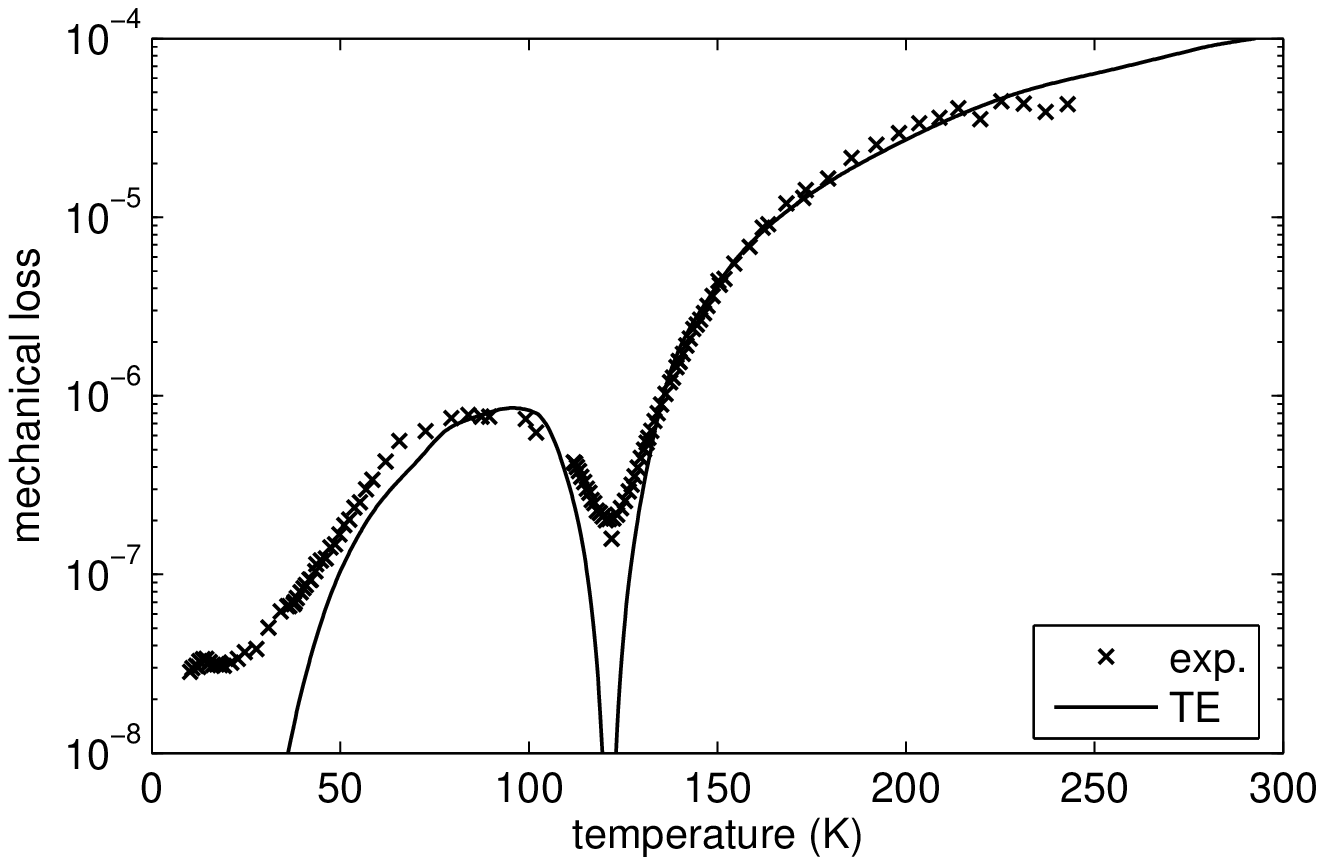}}
\end{center}
\caption{\label{fig:S3}Experimental results of the mechanical loss of sample 3 (dry etching, size).}
%\end{indented}
\end{figure}

Figure\,\ref{fig:S1} summarizes the results obtained for the sample 1. The loss of all modes is thermoelastically limited at temperatures above 150\,K. Below that temperature the modes show an excess loss compared to the predicted thermoelastic limit. The 19.98\,kHz mode shows an onset of the thermoelastic dip at 125\,K and reaches the lowest loss at the lowest temperature of about $5\times10^{-8}$. All other modes had a lowest loss well above $10^{-7}$. A loss peak occured in the results of the 19.98\,kHz mode. This peak was just observed in this specific mode. However, all other modes have shown higher losses which might have covered the loss peak. Alternatively, this peak could have its origin from a resonant coupling to internal modes of the clamping structure. Figure\,\ref{fig:S2} compiles the results for sample 2. In contrast to sample 1 this flexure was fabricated by means of dry etching. The modes follow the thermoelastic limit down to 150\,K. Below that temperature they show evidence of the vanishing thermoelastic damping at around 125\,K. The 3$^{rd}$ bending mode at 1958\,Hz shows an excess loss at temperatures below 125\,K. Here, a coupling between the sample's resonant frequency and the clamp is causing the additional damping. The lowest mechanical loss of $4.2\times10^{-8}$ was observed for the 18.9\,kHz mode again at the lowest temperature. The mode with the highest observed frequency showed some deviations from the predicted thermoelastic loss below 125\,K. The reason for this behavior is the deviation from the pure bending which is the prerequisite for the validity of eq.\,(\ref{eq:TE}). Higher modes show complicated mode shapes including more and more torsional contributions. These contributions do not cause a time varying change in the stress tensor. Pure torsion does not result in a volume change and thus does not create a dissipative heat flux (see eq.\,(\ref{eq:gen_heatcond})). Thus, the amount of thermoleastic damping is smaller than predicted by the simplified equation.
Comparing the results to those of sample 3 (figure\,\ref{fig:S3}) it is obvious that the excess loss seems to be only present at low frequency modes. This was a general observation throughout our investigations. All low frequency modes showed a slightly higher unexpected mechanical loss. Results for low frequency modes have been published previously on wet etched samples \cite{Reid2006}. The results for sample 3 reveal an even lower mechanical loss for this sample reaching $3\times10^{-8}$ at around 10\,K. 

\begin{figure}[b!]
\begin{center}
\subfigure[]{\includegraphics[scale=0.52]{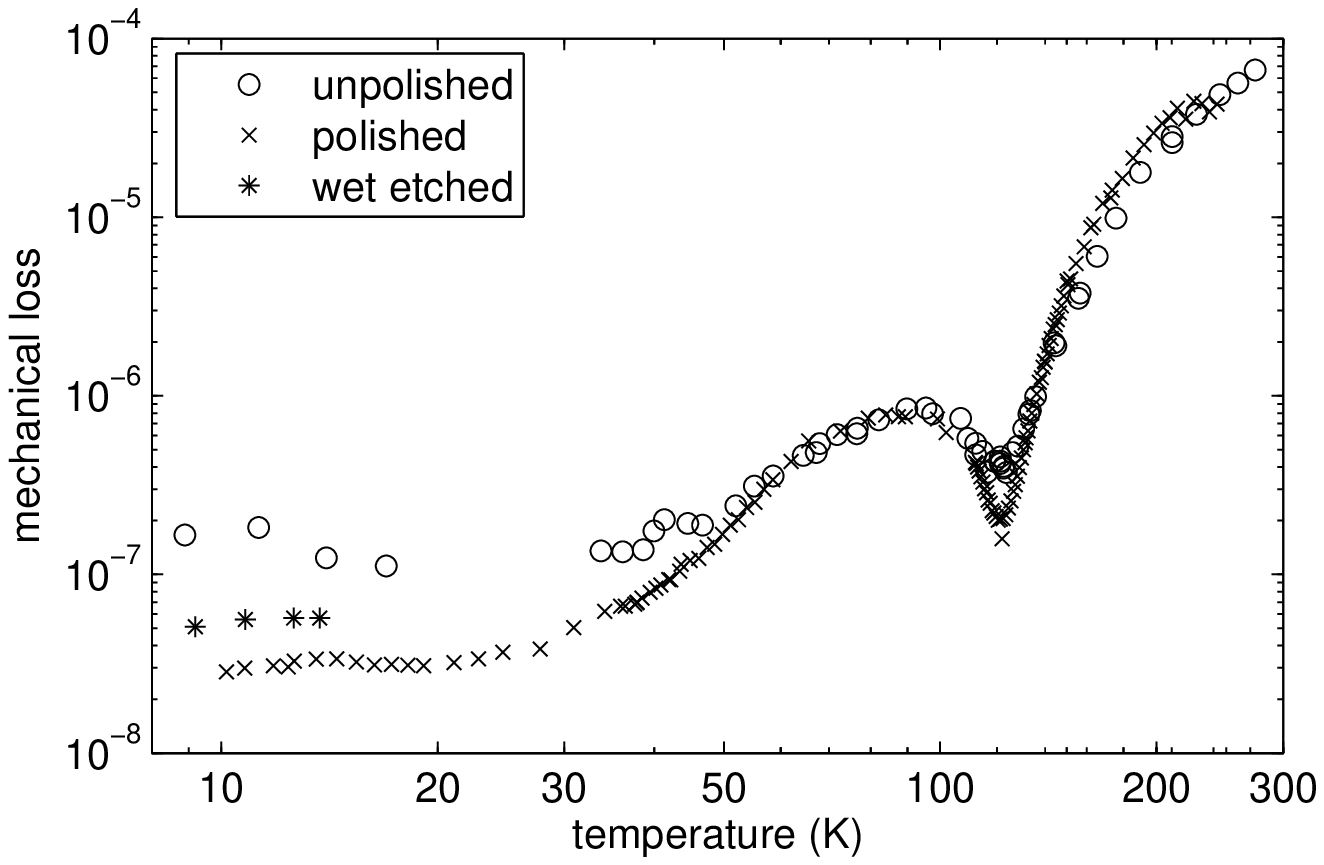}}
\subfigure[]{\includegraphics[scale=0.52]{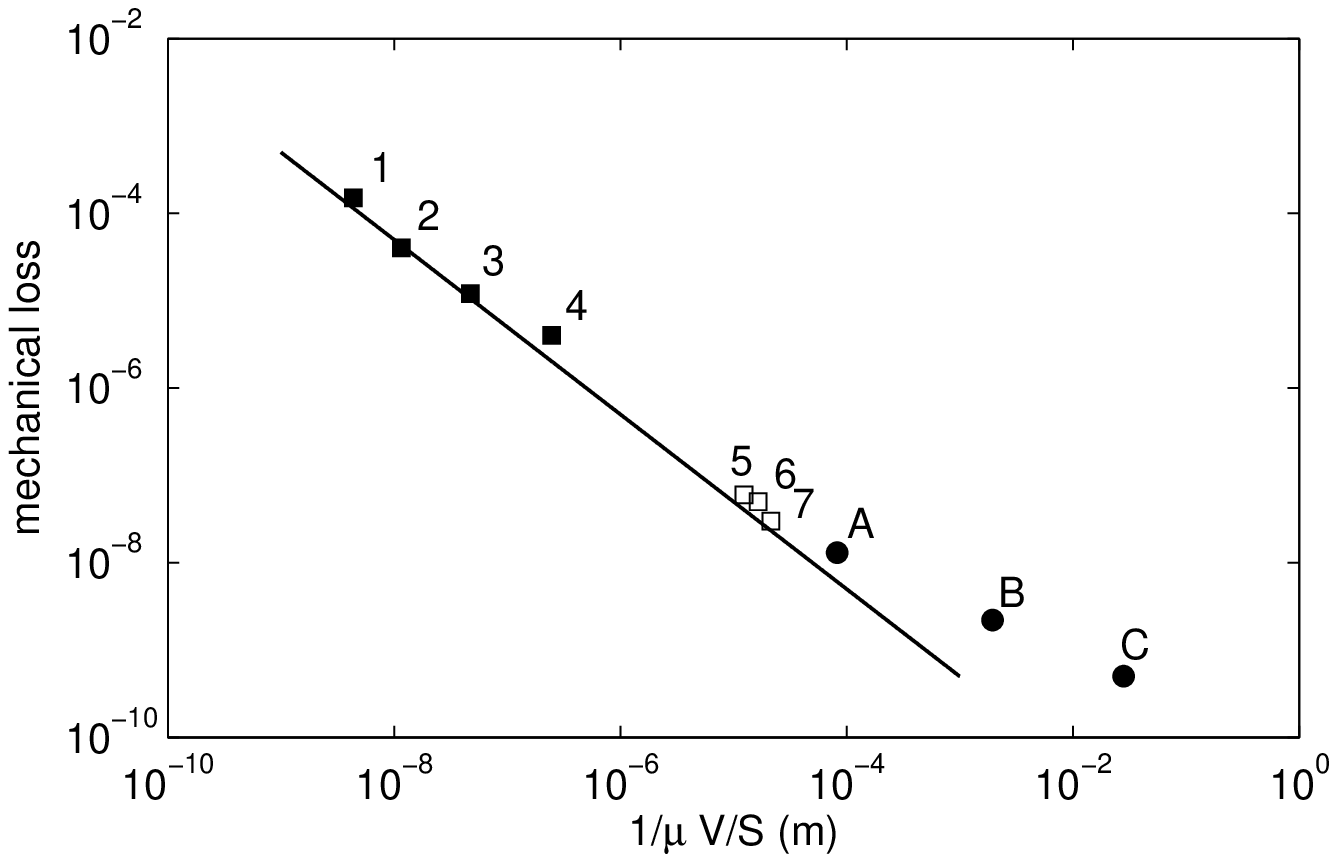}}
\end{center}
\caption{\label{fig:surf}(a) Investigation of the influence of the surface roughness on the mechanical loss. The polished sample corresponds to sample 3 and the unpolished to sample 4 from table \ref{tab:samples}. Both plots are obtained from the 18.4\,kHz mode (6$^{th}$ bending mode). (b) Summary of the lowest loss obtained from different sized silicon oscillators. The numbers and letters are explained in table\,\ref{tab:surf_samples} and the text. The plotted line corresponds to a surface loss parameter $\alpha_s$ of 0.5\,pm.}
\end{figure}

The results suggest that thin samples have slightly higher losses than thicker. This is a first indication of surface dependent losses which predict an increasing loss for thinner samples due to the increased surface-to-volume ratio. However, it needs to be first clarified that the low temperature loss is caused by the surface quality. Therefore, two identical samples with different surface preparation have been measured (sample 3 and 4). While sample 3 had a dry etched and a mechanically polished surface sample 4 consisted of a dry etched and an unpolished (lapped) surface. Sample 3 had two surfaces with a small surface roughness (see table\,\ref{tab:samples}). In contrast, sample 4 had one surface with a 50 times larger roughness. The results are summarized in figure\,\ref{fig:surf}(a). A log-log-plot was chosen to emphasize the low temperature part of the measurement. Both samples show a similar behavior for temperatures above 50\,K. They both follow the thermoelastic limitation. Below 50\,K sample 4 has a lowest mechanical loss above $10^{-7}$ whereas sample 3 showed a minimum loss of $3\times10^{-8}$. Both measurements have been obtained at similar conditions: 10\,K and the 18.4\,kHz mode (6$^{th}$ bending mode). This measurement is evidence that the surface of the sample, rather than the setup, is limiting the mechanical loss at low temperatures. 

The wet etched sample had a different geometry and thus it is not possible to compare the results at the same frequency and modeshape. However, it is possible to compare the results for the 19.98\,kHz mode of the wet etched sample with the dry etched samples although it is not the same modeshape. Due to the observed loss peak with unknown origin (see figure\,\ref{fig:S1}(c)) only the lowest measured losses at temperatures below 14\,K were plotted in figure\,\ref{fig:surf}(a) for comparison. The obtained loss lies in between sample 3 and 4. This correlates with the roughness of the wet etched sample of 33\,nm which is in between the values for the other samples. The results suggest a correlation between the roughness of the sample and the mechanical loss obtained at low temperatures.

\subsection{Analysis of the surface loss}
\label{sec:surf_loss}

In order to extract a value for the surface loss of silicon at low temperatures the surface loss model by Gretarsson and Harry (see eq.\,(\ref{eq:surf_loss_simple})) was applied to available data. Eq.\,(\ref{eq:surf_loss_simple}) suggests a linear correlation between the mechanical loss and $\mu S/V$. Figure\,\ref{fig:surf}(b) summarizes different mechanical losses obtained in several experiments. Only long time experiments at low temperatures have been included in this analysis due to their applicability to a use of silicon in GW detectors or other long time experiments. It is well known that a lower surface loss can be obtained for special treatments of the surface like heating up to 1000 degrees for several seconds \cite{Yang2000,Ono2003,Liu2005}. However, this lower surface loss disappears after some seconds or minutes due to a possible surface contamination that covers the surface after that time. Due to the expected operational time of hours, days or even months and years for long time experiments like the GW detection these values have not been included into our consideration. Additionally, only values from the literature have been chosen where it was possible to extract the geometry and the modeshape to calculate the geometry factor $\mu$. Only experiments have been selected on silicon flexures at low temperatures (T $\leq$ 10\,K) to be able to neglect thermoelastic damping. The numbers at the plot indicate the original reference for the values. All key parameters of the different experiments are compiled in table\,\ref{tab:surf_samples}.

\begin{table}
\caption{\label{tab:surf_samples}Summary of the parameters of the oscillators from figure\,\ref{fig:surf}(b) used to obtain the surface loss parameter from literature values.}
\begin{indented}
\item[]\begin{tabular}{@{}llllll}
\br
point & T (K) & $\phi$ & $\mu$ & geometry & reference\\
\mr
1 & 4.8 & $1.5\times10^{-4}$ & 2.98 & $220\,\mu m \times 5\,\mu m \times 0.06\,\mu m$ & \cite{Stowe1997}\\
2 & 4.2 & $4.0\times10^{-5}$ & 2.99 & $300\,\mu m \times 10\,\mu m \times 0.07\,\mu m$ & \cite{Yasumura2000}\\
3 & 4.2 & $1.2\times10^{-5}$ & 2.86 & $260\,\mu m \times 3.9\,\mu m \times 0.29\,\mu m$ & \cite{Mamin2001}\\
4 & 6   & $5.0\times10^{-6}$ & 2.94 & $470\,\mu m \times 45\,\mu m \times 1.5\,\mu m$ & \cite{Wago1996}\\
\mr
5 & 7   & $5.5\times10^{-8}$ & 2.96 & sample 1 & \\
6 & 10  & $5.0\times10^{-8}$ & 2.96 & sample 2 & \\ 
7 & 10  & $3.0\times10^{-8}$ & 2.97 & sample 3 & \\ 
\mr
A & 7   & $1.2\times10^{-8}$ & 2.98 & $\oslash100\,mm \times 0.5\,mm$ & \cite{Zendri2008}\\
B & 5   & $2.2\times10^{-9}$ & 2.36 & $\oslash76\,mm \times 12\,mm$ & \cite{Nawrodt2008}\\
C & 2   & $5.0\times10^{-10}$ & 0.77 & $\oslash106\,mm \times 229\,mm$ & \cite{McGuigan1978}\\
\br
\end{tabular}
\end{indented}
\end{table}

Plotting the mechanical loss of a silicon flexure based oscillator against its volume-to-surface ratio $V/S$ results in a linear dependency if the mechanical loss is determined by surface loss. The mode shape dependent factor $\mu$ was obtained from eq.\,(\ref{eq:mu_simple}) for all Si flexures (1-7). The geometry factor for the bulk samples (A-C) has been obtained from a numerical derivation of eq.\,(\ref{eq:mu}) using the FEA software COMSOL. $\mu$ is close to 3 for all cantilevers and the thinnest bulk sample. The thick samples show a deviation which is largest for sample C. Here, the thickness is much larger than the diameter. This causes only a small fraction of the test body vibration to probe the surface loss. The effect of the surface loss to this specific mode is thus reduced.

Eq.\,(\ref{eq:surf_loss_simple}) has been used to model the data obtained for the thin flexures (1-7). It was assumed that the intrinsic bulk loss is much smaller than the surface loss. The only free parameter in eq.\,(\ref{eq:surf_loss_simple}) is the surface loss parameter $\alpha_s$ under this assumption. It was adjusted in a way that the resulting line corresponds well to the lowest losses of the silicon flexures. This allows a rough estimate of the surface loss parameter of 0.5\,pm with an error of about 25\%. 

The mechanical loss values obtained from the flexures follow the predictions of the surface loss model. The bulk samples show a deviation towards higher losses. Here, the neglected intrinsic bulk loss starts to become significant. It is possible to estimate an averaged surface loss $\phi_{surf}$ assuming a homogeneous surface layer of thickness $t_s$ and a Young's modulus equal to the bulk value. Eq.\,(\ref{eq:dissdepth}) gives then:

\begin{equation}
d_s = t_s \frac{\phi_{surf}}{\phi_{bulk}}.
\end{equation}

\noindent This leads to a value of 

\begin{equation}
\phi_{surface} = \alpha_s/t_s
\end{equation}

\noindent for the surface loss. The thickness of the lossy layer $t_s$ is unknown and can only be approximated. Assuming similar values compared to silica ($t_s \approx 1\,\mu m$, \cite{Gretarsson1999}) gives a homogeneous surface loss value of $5\times10^{-7}$ which is a factor of 20 smaller than for fused silica \cite{Gretarsson1999}. However, this estimate is based on the weak assumption of a similar thickness of the lossy surface layer in silicon and silica.

\section{Conclusions}

We presented measurements of the mechanical loss of silicon flexures in a temperature range from 5 to 300\,K and frequencies from 1 to 86\,kHz. At temperatures above 100\,K the experimental data follows the thermoelastic predictions. The observed loss is higher than the thermoelastic predictions for temperatures below 50\,K. The level of the mechanical loss at low temperatures is strongly dependent on the surface quality. Rough surfaces produce a higher mechanical loss than etched smooth surfaces. The lowest mechanical loss of $3\times10^{-8}$ was obtained for a 130\,$\mu$m thick cantilever at a frequency of 18.4\,kHz and a temperature of 10\,K. Using the lowest mechanical loss obtained from different silicon based oscillators it was possible to extract a surface loss parameter according to the surface loss model from Gretarsson and Harry of $\alpha_s = 0.5\,pm$. 
This value is at least an order of magnitude smaller than for fused silica. Together with its remarkable mechanical and thermal properties the small surface loss makes silicon a very promising material for low mechanical loss applications at cryogenics like suspension elements in future gravitational wave detectors. However, the origin of the surface loss is still not fully understood and will be within the focus of further investigations.

\ack
\small

This work was in part supported by the German Science Foundation under contract SFB TR7 and the STFC in the UK. R. Nawrodt is supported by the FP7 EU project Einstein Telescope Design Study under contract number 211743. I. W. Martin holds an STFC postdoctoral fellowship. S. Reid holds a Royal Society of Edinburgh Research Fellowship.
The authors would like to thank B. H\"ofer from the Fraunhofer Institute f\"ur Angewandte Optik und Feinmechanik for support in the sample preparation. We would like to thank P. Hanse and S. Eiweleit for their support constructing the setup. The help of L. F\"ollmer and T. Ei{\ss}mann is kindly acknowledged during the extended 24\,h cryogenic runs.

\section*{References}

\bibliographystyle{unsrt}      % basic style, author-year citations
\bibliography{Si_flex}

%\begin{thebibliography}{10}
%\bibitem{X} Goosens M, Rahtz S and Mittelbach F 1997 {\it The \LaTeX\ Graphics Companion\/} 
%(Reading, MA: Addison-Wesley)
%\bibitem{Rowan2005} Rowan S, 
%\bibitem{eps} Reckdahl K 1997 {\it Using Imported Graphics in \LaTeX\ } (search CTAN for the file `epslatex.pdf')
%\end{thebibliography}

\end{document}